\input harvmac
\newcount\figno
\figno=0
\def\fig#1#2#3{
\par\begingroup\parindent=0pt\leftskip=1cm\rightskip=1cm\parindent=0pt
\baselineskip=11pt
\global\advance\figno by 1
\midinsert
\epsfxsize=#3
\centerline{\epsfbox{#2}}
\vskip 12pt
{\bf Fig. \the\figno:} #1\par
\endinsert\endgroup\par
}
\def\figlabel#1{\xdef#1{\the\figno}}
\def\encadremath#1{\vbox{\hrule\hbox{\vrule\kern8pt\vbox{\kern8pt
\hbox{$\displaystyle #1$}\kern8pt}
\kern8pt\vrule}\hrule}}

\overfullrule=0pt

\Title{MIT-CTP-2627}
{\vbox{\centerline{Absorption of angular momentum 
 by black holes 
and D-branes}}}
\smallskip
\smallskip
\centerline{Samir D. Mathur\foot{E-mail: me@ctpdown.mit.edu}}
\smallskip
\centerline{\it Center for Theoretical Physics}
\centerline{\it Massachussetts Institute of Technology}
\centerline{\it Cambridge, MA 02139, USA}
\bigskip

\medskip

\noindent
We consider the absorption of higher angular momentum modes of scalars 
into black holes, at low energies, and ask if the resulting cross sections 
are reproduced by a D-brane model.  To get the correct dependence on the
 volume of the compactified dimensions,  we must let the absorbing 
element in the brane model have  a tension that is the geometric mean 
of the tensions of the D-string and an effective stringlike tension obtained
 from the D-5-brane; this choice  is
 also motivated by T-duality. In a dual model we note that the correct
 dependence on the volume of the compact dimensions and the coupling
 arise if the absorbing string is allowed to split into many strings
 in the process of absorbing a higher angular momentum wave.
We obtain the required energy dependence 
of the cross section by carrying out the  integrals resulting from
 partitioning the
energy of the incoming quantum into vibrations of the string.

\Date{April, 1997}

\def\LS{{L^{(S)}}}
\def\TD{{T^{(D)}}}
\def\TS{{T^{(S)}}}

\def\TDF{{{T_5}^{(D)}}}

\def\np {{  Nucl. Phys. }}
\def \pl {{  Phys. Lett. }}

\newsec{Introduction.}

With the development of string theory and the ideas of  duality,  there has
 been considerable progress in
our understanding of black holes. Following suggestions of Susskind, 
the number of string theory states at weak coupling have been found to agree
 with the number of states expected from the Bekenstein entropy of the hole
 that would form at strong coupling
\ref\suss{L. Susskind, hep-th/9309145;
J. Russo and L. Susskind, {\it Nucl. Phys.}~{\bf B437} (1995) 611.}\ref\sen{
A. Sen,
{\it Nucl. Phys.}~{\bf B440} (1995) 421 and {\it Mod. Phys. Lett.}
~{\bf A10} (1995) 2081.}\ref\stromvafa{ A. Strominger and C. Vafa, 
\pl B379 (1996) 99, hep-th/9601029.}\ref\callanmalda{C.G. Callan and
 J.M. Maldacena, \np B472 (1996) 591, hep-th/9602043.}.
Further, it was found that the rates of absorption and emission of 
minimal scalars computed at weak coupling matched the Hawking radiation
 rates expected from the black hole at strong coupling 
\ref\dasmathurone{S.R. Das and S.D. Mathur,
Nucl. Phys. {\bf B478} (1996) 561, hep-th/9606185. }
\ref\dasmathurtwo{S.R. Das and S.D. Mathur, Nucl. Phys.
 {\bf B482} (1996) 153, hepth 9607149.}.
This  calculation has been extended to emission of charged quanta
\ref\gubserklebanov{S.S. Gubser and I.R. Klebanov, 
Nucl. Phys. B482 (1996) 173, hep-th/9608108.}, to  higher orders in
 the energy of the incident quantum under certain conditions 
\ref\maldastrom{J.M. Maldacena and A. Strominger, Rutgers preprint RU-96-78, 
hep-th/9609026.}, and to nonminimal `fixed' scalars \ref\cgkt{C.G. Callan, 
Jr., S.S. Gubser, I.R. Klebanov and A.A. Tseytlin,
hep-th/9610172.}.

To make contact with the black hole information paradox, we need to understand 
the absorption of quanta that are small in size compared to the horizon, so 
that they can fall into the horizon through a reasonably localised direction.
 This implies that we understand the absorption of higher angular momentum
 modes, since a wavepacket that is to be localised in the angular directions 
must be composed of several components of angular momentum $l$. It is also 
important to understand
the absorption at wavelengths small compared to the size of the horizon, since
 that too is required to well localise the infalling quantum.

In this paper we discuss the absorption of low energy higher $l$ modes for 
minimal scalars, by the classical black hole and by the D-brane model of the
 hole. For the case $l=0$ it had turned out to be adequate to use a model
 where a D-string absorbed an incoming scalar by converting its energy  into
  vibration modes on the D-string \dasmathurone . An equivalent result was 
obtained in the S-dual model where the absorbing element was an elementary
 string, and
the absorption amplitude was computed using standard perturbative string 
theory \dasmathurtwo .

But there are difficulties with naively extending these models to the absorption
 of quanta with higher $l$. Let us consider the model of the 4+1 dimensional black
 hole introduced in \callanmalda . The spacetime has total dimension 10, of which 
5 space dimensions are compactified on a 5-torus $T^5=T^4\times S^1$. The $S^1$ 
is the direction in which the absorbing string is wound, while the 5-branes in 
the model wrap all over the $T^5$. Let the volume of the $T^4$ be $V_4$, and 
the length of the $S^1$ be $L$. 

The incident quantum is expected to convert its energy into modes that travel
 in the direction of the circle $S^1$. As the angular momentum $l$ of the 
incident quantum is increased, we expect that more and more such (fermionic)
 modes will be created. But if these modes are vibrations of the D-string, 
then the absorption calculation will be confined to the vicinity of this string, 
and will not be sensitive to the volume $V_4$ that is available transverse to 
the D-string. Thus the $V_4$ dependence of the cross section will not change with $l$.
But the classical cross section {\it does} depend on $V_4$; this dependence is 
 $\sim V_4^{-1-l}$.

If we use a dual model where  the absorbing element is an elementary string, 
and consider the absorption process as a fundamental string interaction diagram,
 then we see that the incoming massless
quantum can create no more than 4 new fermions in the final state,
if we use the three point tree vertex. This is because the world sheet conformal 
theory is a free theory, and the massless scalar has at most two fermionic 
oscillator to contribute to each of the left and right sides. But we need a
 number of fermions that increases without bound with increasing $l$.
If we allow loops, then we get additional powers of $g^4$ in the cross section
 for every extra loop, while the classical cross section is seen to increase
 by powers of $g^2$ as $l$ increases by one unit.

In this paper we do the following:

(a)\quad We observe that in the D-brane model we get the correct dependence of
 the absorption cross section on $V_4$ if we let the absorbing element be a
 long string with a tension that is the geometric mean of the tension of the
 D-string and the effective tension obtained for vibrations of the 5-D-brane
 which travels in the long direction $S^1$. We note that such a choice is also 
T-duality symmetric.

We also note that in the dual elementary string model, the correct dependence 
on $V_4$ and $g$ is obtained if we allow the initial string to split into $l+1$
 strings when absorbing a quantum of angular momentum $l$. The details of the
 amplitude calculation are
however  not very clear in such a model.

(b)\quad The energy of the incoming quantum is expected to be shared between
 a pair of bosonic quanta and $2l$ fermionic quanta,
travelling in the direction of the circle $S^1$. We carry out the integrals 
over momenta, and obtain the energy dependence that is required by the classical
 cross section.

Recently the absorption of higher angular modes has been considered for 3-branes
\ref\klebanovthree{S.S, Gubser, I. Klebanov and A.A.  Tseytlin, hepth 9703040.}
 and for the 4+1 dimensional black hole with three charges
through an effective conformal theory 
\ref\maldastromtwo{J. Maldacena and A. Strominger, hepth 9702015.}.
 There now exist a large number of results pertaining to the 
black hole - D-brane comparison.
The behavior of the D-brane as a black body was discussed in
\ref\dharetal{A. Dhar, G. Mandal and S.R. Wadia, hepth 9605234.},
 where it was shown that emissions of quanta are proportional to
the classically expected emissions.
The issue of higher orders in coupling was discussed in \ref\das{S.R. Das,
hepth 9703146.}.  Comparisons of brane and  classical absorption were
 discussed in
\ref\jenny{F. Dowker, D. Kastor and J. Traschen, hepth 9702109.}
\ref\dealwis{S.P. de Alwis and K. Sato, hepth 9611189.}. 

The plan of this paper is as follows. In section 2 we discuss the classical
 cross section. In section 3 we discuss the issue of $V_4$ dependence in the 
D-brane model. Section 4 discusses dependence on other parameters of the model.
In section 5 we discuss a possible description in the dual model where we 
 use the elementary string Polyakov amplitudes. In section 6 we discuss the energy 
dependence of
the amplitudes. Section 7 is a discussion.

\newsec{The classical absorption cross section.}

The metric of the 5-dimensional hole is
\eqn\onenten{
   d s_{5}^2 = -f^{-2/3} h d t^2 + 
                 f^{1/3} 
                  \left( {h\inv} d r^2 + r^2 d \Omega_{3}^2 \right)
 \ , }
where
\eqn\onentenp{ h(r)= \left (1 - {r_0^2\over r^2}\right )\ , \qquad 
f(r)= \left (1 + {r_5^2\over r^2}\right )
\left (1 + {r_1^2\over r^2}\right ) \left (1 + {r_p^2\over r^2}\right )
\ .
}

Let
\eqn\onentenpp{r_p^2=r_0^2\sinh^2\sigma_p }
We will be in the region of parameter space where $r_p\sim r_0 << r_1, r_5$. Thus
only the momentum-antimomentum excitations on the string are excited; the
 excitations of strings-antistrings and 5-branes-anti 5-branes is suppressed.

We will consider the absorption of a graviton that is a scalar 
from the 5-dimensional point of view. 
This is a minimally coupled scalar, and satisfies the free wave equation 
on the 5-dimensional black hole metric. The absorption probability for 
a spherical wave
of angular momentum $l$ was computed in 
\ref\klebanovmathur{I. Klebanov and S.D. Mathur, hepth 9701187.}\maldastromtwo .
 In the limit
where $r_1, r_5>>r_0, r_p$ this probability is
\eqn\onenone{a^l={\omega^3\over 4\pi}{A_H\over [l!(l+1)!]^2}
({\omega r_0\over 2})^{2l}
|{\Gamma({(l+2)\over 2} -i{\omega\over 4\pi T_L})
\Gamma({(l+2)\over 2} -i{\omega\over 4\pi T_R }) \over
 \Gamma(1-i{\omega\over 2\pi T_H })}|^2  }
where 
\eqn\onentwo{A_H=2\pi^2r_1r_5r_p\cosh\sigma_p}
is the area of the horizon. The temperature of the black hole is
\eqn\onenthree{T_H={r_0\over 2\pi r_1r_5\cosh\sigma_p} }
The left and right temperatures are \maldastrom\
\eqn\onenfour{T_L={r_0e^{\sigma_p}\over 2\pi r_1r_5}, ~~T_R=
{r_0e^{-\sigma_p}\over 2\pi r_1r_5}}
The absorption cross  section for angular momentum $l$ is
(see Appendix C for details)
\eqn\onenfive{\sigma^l=(l+1)^2{4\pi\over \omega^3}a_l }

We have, for $l$ odd
\eqn\oneonez{\eqalign{\sigma^l=(l+1)^2{\pi^3\over 2^{4l}}&
{(r_1r_5)^{2l+2}\omega^{2l-1}\over [l!(l+1)!]^2}
\cr
[\omega^2+&(2\pi T_L)^21^2][\omega^2+(2\pi T_L)^23^2]\dots
 [\omega^2+(2\pi T_L)^2l^2]\cr
[\omega^2+&(2\pi T_R)^21^2][\omega^2+(2\pi T_R)^23^2]\dots
 [\omega^2+(2\pi T_R)^2l^2]\cr
&{e^{\omega/T_H}-1\over (e^{{\omega\over 2T_R}}+1)
(e^{{\omega\over 2T_L}}+1)}\cr} } 

For $l$ even
\eqn\oneonezp{\eqalign{\sigma^l=(l+1)^2{\pi^3\over 2^{4l}}&
{(r_1r_5)^{2l+2}\omega^{2l+1}\over [l!(l+1)!]^2}
\cr
[\omega^2+&(2\pi T_L)^2 2^2][\omega^2+(2\pi T_L)^2 4^2]\dots
 [\omega^2+(2\pi T_L)^2l^2]\cr
[\omega^2+&(2\pi T_R)^2 2^2][\omega^2+(2\pi T_R)^2 4^2]\dots
 [\omega^2+(2\pi T_R)^2l^2]\cr
&{e^{\omega/T_H}-1\over (e^{{\omega\over 2T_R}}-1)
(e^{{\omega\over 2T_L}}-1)}\cr} }

As a check we note that for $l=0$, $\omega\rightarrow 0$, we 
obtain $\sigma=A_H$
in accordance with the universal form of the low energy cross 
section for minimal scalars
\ref\dgm{S. Das, G. Gibbons and S.D. Mathur, hep-th/9609052.}.

\newsec{$V_4$ dependence.}

\subsec{The behavior of the classical cross section.}

In terms of microscopic variables, we have for the D-brane model
 \ref\hms{G. Horowitz,
 J. Maldacena and A. Strominger, Phys. Lett {\bf B383} (1996) 151,
hep-th/9603109.}:
\eqn\onethree{r_1=[{g n_1\over (2\pi)^2 V_4}]^{1/2}\LS^3 }
\eqn\onefour{r_5=[{g n_5\over (2\pi)^2}]^{1/2} \LS  }
Here $n_1$, $n_5$ are the numbers of D-strings and D-5-branes respectively. 
$V_4$ is
 the volume of the 4-torus transverse to the direction in which the D-string
 is wound.
 $\LS$ is the string length, defined so that under T-duality, a circle of
 circumference
 $A\LS$
goes to a circle of circumference $A^{-1}\LS$. $g=e^\phi$ is the elementary
 string 
coupling. The tension of the elementary string is $\TS=2\pi\LS^{-2}$. 
The tension 
of the D-string is
\eqn\twonone{\TD=g^{-1}\TS=2\pi\LS^{-2}g^{-1} }
The tension of the 5-D-brane is
\eqn\twontwo{\TDF=g^{-1}\TS\LS^{-4}=2\pi\LS^{-6}g^{-1} }

The cross section \oneonez , \oneonezp\ is seen to depend on $V_4$ as
\eqn\twonthree{\sigma^l\sim V_4^{-(l+1)} }
(The product $r_1r_5$ depends on $V_4$ as $\sim V_4^{-1/2}$.)

Suppose we have a bound state of 5-D-branes and D-strings.
In the effective string model of \ref\maldasuss{J.M. Maldacena and L. Susskind, 
Stanford preprint
SU-ITP-96-12, hep-th/9604042.}
  the effect of the 5-D-branes can be taken into account through
a fractionation of the tension of the D-string:
\eqn\twonfour{T_{eff}=\TD n_5^{-1} }
This model was motivated by performing a duality on the case studied in
 \ref\dasmathur{S.R. Das and S.D. Mathur, Phys. Lett. B375 (1996) 103, 
 hep-th/9601152.}.
  In the latter case it was shown by using S-duality that the momentum
 modes that travel on a D-string can, under some conditions, be `fractional'
 (i.e. go in units of
$2\pi/(n_1L)$ rather than units of $2\pi/L$), though the total momentum on 
the collection of strings must be still quantised in interger units 
(i.e. must be an integer multiple of $2\pi/L$).
A sequence of dualities can map the D-string to a 5-D-brane, and the
 momentum mode to a D-string bound to the 5-D-brane. Then it is
a plausible conjecture that at least for some dynamical purposes the
 D-strings bound to D-5-branes should be `fractional', with tension $\TD/n_5$.

Now suppose we compute the absorption of a minimally coupled scalar
by the D-string. If the only effect of the 5-D-branes comes through
\twonfour , then we see that the physics of oscillations on the D-string 
is not sensitive to the volume $V_4$. In the absorption calculation, the
 only way that $V_4$ will enter will be through the
normalisation factor for the incoming scalar, which propagates in the full
 10-dimensional spacetime and has in its normalisation a
factor $V_4^{-1/2}$. This would yield $V_4^{-1}$ in the cross section, which 
is indeed appropriate to the $l=0$ cross section: 
\eqn\twofive{\sigma^0=A_H=4G^{(5)}S_{Bek}=8\pi\sqrt{n_1n_5n_p}
G^{(5)}\sim V_4^{-1} }
where we have noted that
\eqn\twoone{G^{(5)}=G^{(10)}V_4^{-1}L^{-1} \sim V_4^{-1}}

But for the case of $l>0$ we would continue to find the  $V_4^{-1}$ dependence,
 since the quanta created on the D-string would not see
the size of the transverse $T^4$. This is in contradiction with
\twonthree . 

\subsec{Duality considerations, and the `mean string'.}

Let us take one 5-D-brane bound to one D-string.
Consider a situation where the length $L$ of the circle where the D-string is
 wrapped is very long compared to the sides of the compact torus perpendicular
 to the D-string, which we take to be of order $V_4^{-1/4}$ each. The 5-brane 
is wrapped on $T^4\times S^1$, so it also sees the length $L$. Further, in this
 case we can consider excitations of wavelength
\eqn\twonfive{V_4^{-1/4}<<\lambda\sim L}

There are two kinds of such excitations that we can naively see in this system.
 If we oscillate the D-string, we get vibrations on a string with tension:
\eqn\twonsix{T_1=\TD}
If we oscillate the 5-D-brane, then we expect that this will behave as a string
 with tension
\eqn\twonseven{T_2=V_4\TDF=\TD{V_4\over \LS^4} }

If we perform four T-dualities, in the four directions of the torus $T^4$, then
 the D-string will become a 5-D-brane, and the 5-D-brane will become a D-string.
 The new coupling will be
\eqn\twoneight{g'=g[{V_4\over \LS^4}]^{-1} }
and the new volume of $T^4$ will be
\eqn\twonnine{V_4'=[{V_4\over \LS^4}]^{-1} \LS^4 }

The 5-D-brane becomes a D-string, which has the tension
\eqn\twonten{\TD'\equiv T_1'=\TS (g')^{-1}=\TD [{V_4\over \LS^4}]}
which agrees with \twonseven\ as it should. The original D-string becomes a 5-D-brane
 with tension
\eqn\twonel{\TDF'=\TS (g')^{-1}\LS^4=\TD [{V_4\over \LS^4}]\LS^{-4} }  
so the effective tension for the long wavelength modes considered here is
\eqn\twontw{T_2'=\TDF' V_4'=\TD}
Thus the two tensions $T_1, T_2$ get interchanged under T-duality.
From the point of view of the noncompact spacetime, both  tensions $T_1, T_2$ that
 appear are on a symmetrical footing.
Thus it is not natural to choose either as the effective tension for the vibrations 
that are excited on the system. We take instead the geometric mean of the two tensions
\eqn\twonth{T_m=\sqrt{T_1T_2}=\TD [{V_4\over \LS^4}]^{1/2} }
We will have $4n_1n_5$ bosonic degrees of freedom on this string, together with
 their $4n_1n_5$ fermionic superpartners.
Let us call the string with this tension the `mean' string to 
differentiate it from the string with tension given by \twonfour ,
which is usually termed the `effective string'.

We can still have the case that the $\sim n_1n_5$
degrees of freedom in certain domain of parameters give 4 bosonic and 4 fermionic
 degrees  of freedom on a circle of length $n_1n_5L$, just as was the case for
 the effective string model. But note that the tension $T_m$ does not give
in any simple way the mass of either the D-strings or the 5-D-branes in the 
system. It is an effective parameter for the excitations of the D-string - 5-D-brane 
bound state.

\subsec{Disc diagram calculations.}

When the incoming scalar is absorbed in the brane model, we expect that there is
one bosonic excitation created on each of the left and right sides; these bosons 
carry the
spins of the scalar. there are also $l$ fermions on each side, for absorption of
 angular
momentum $l$. (Some details of the group theoretic structure of partial waves is
 given in Appendix B. 
The above kinds of excitations were also involved in the effective conformal
theory description used in \maldastromtwo .)

Let us see what a calculation using open string disc diagrams would look like, if
 we wish to obtain the $V_4$ dependence required by the classical absorption cross 
section. In the absorption of a $l=0$ minimal scalar by a D-string we had one power 
of $g_{closed}$ from the scalar at the center of the disc, two powers of $g_{open}$ 
from the two open strings created on the D-string, and two negative powers of 
$g_{open}$ from the disc amplitude itself. Thus we were left only with $g_{closed}$
 in the amplitude. Since
$g_{open}^2\sim g_{closed}$, there was however no real significance to separating
 the powers to $g_{open}$ and $g_{closed}$ in this way.

But if we are computing the absorption of $l>0$ partial waves, then there are $2l+2$
 open strings on the disc boundary, besides the closed string at the center.
 We obtain the required $V_4$ dependence by using instead of $g_{open}$ the coupling
\eqn\twonfourt{g^m_{open}= g_{open}[{V_4\over \LS^4}]^{-1/4} }
while leaving $g_{closed}$ the same as before. This choice corresponds to the 
string tension \twonth\ in the same way that we have the usual correspondence
 $\TD\sim g_{open}^{-2}$.

Equivalently, we can still use $g_{open}$ as the coupling but
alter the normalisation factors for the vibrations that are created. Naively
 we have two kinds of vibrations: one where we produce open strings attached 
to the D-string (1-1 strings) and one where we have open strings attached to 
the 5-brane (5-5 strings).
The latter will be taken to have momentum only along the long direction $S^1$.
 But these 5-5 strings will still have a normalisation factor $\sim V_4^{-1/2}$,
 which the 1-1 strings did not have. Since these two kinds of open strings are 
interchanged under T-duality, we would  not know a priori which to use. 
(Note that 1-5 strings go to themselves under the T-duality considered here.)

Again following the path of taking geometric means, we take an effective 
transverse volume of the interaction region   equal to 
$[{V_4\over \LS^4}]^{1/2}\LS^4$. This gives normalisation
 factors for each open string equal to
\eqn\twonfift{{\cal N}\sim [{V_4\over \LS^4}]^{-1/4}\LS^{-2} }
Again note that the $l=0$ case is not altered, since the change 
in the normalisation of the two open strings is compensated by 
the change in the volume of the interaction region, which also 
appears in the amplitude. 

\subsec{Obtaining the $V_4$ dependence.}

In any of the above ways of taking into account the effective tension for
 the vibrations, we get the desired $V_4$ dependence.
The power of $g_{open}$ is ${2l+2-2}=2l$ in the amplitude, $4l$
in the cross section, so that we get an additional factor $V_4^{-l}$ in the
 cross section apart from the $V_4^{-1}$ that arises from the normalisation
 of the incoming scalar.

Equivalently, by using the proposed change of the volume of the interaction
 region, we get a factor $[V_4^{-1/4}]^{2l+2}V_4^{1/2}$
in the amplitude, where we have used the fact that there are $2l+2$ open 
strings and that there is one factor of the volume of the interaction region.
 This gives $V_4^{-l}$ in the cross section, 
again apart from the $V_4^{-1}$ that arises from the normalisation of the 
incoming scalar.

Thus in each case we get the desired $V_4$ dependence.

\newsec{$g$, $n_1$, $n_5$, $L$ dependence.}

In the classical cross section, we note that with regard to $g$, $L$ dependence,
\eqn\twopptwo{r_1\sim [G^{(5)}L\TS g^{-1}]^{1/2}\sim g^{1/2} }
\eqn\twoppthree{r_5\sim [G^{(5)}V_4L\TS^3 g^{-1}]^{1/2} \sim g^{1/2}}
\eqn\twoppfour{r_1r_5\sim g }

Further with regard to the dependence on the number of 1-branes and 5-branes,
\eqn\twoppfive{(r_1r_5)^2\sim n_1n_5, ~~~\sigma^l\sim (n_1n_5)^{l+1}}
Thus
\eqn\twonnone{\sigma^l\sim g^{2l+2}(n_1n_5)^{l+1} }

All these dependences are seen to result if we assume that we have $\sim n_1n_5$ 
degrees of freedom on a very long string. The local nature of the interaction says
 that the cross section 
is not sensitive to the length $L$ of the string, apart from the $l$ independent
 factors that were found in \dasmathurtwo\ in the case for $l=0$. A disc diagram
 with $2l+2$  1-5 open strings 
gives a cross section that goes like
$(n_1n_5)^{l+1}$ from the sum over flavours of the open strings, after we note that
1-5 and 5-1 open strings alternate around the disc boundary, with flavours
agreeing at the junctions where two open strings meet.
Further the disc amplitude goes like $g_{open}^{2l+2}g_{closed}g_{open}^{-2}$,
which gives $\sim g_{closed}^{2l+2}$ in the cross section. (These
dependences on $g$ and $n_1n_5$ were  noted in \maldastromtwo .)

Thus note that here we seem to need that the different strands of the string
 interact locally with each other, and the essential physics  is not
  contained in just the vibrations of one long effective string.

\newsec{The dual model.}

In \dasmathurtwo\ we had seen that the leading term in the absorption of minimal scalars
could be reproduced from a calculation where the absorption of the incident scalar by
the string present in the black hole model was viewed as a three point vertex
of ordinary string theory. For convenience we use the S-dual model to the brane
model used in the preceeding sections, though the same method could be applied to either
model with suitable changes of string tensions and couplings. We wish to see if
some fundamental string interaction vertex reproduces the $V_4$ and $g$
dependences required by the classical cross section.

\subsec{$V_4$, $g$ dependence.}

We consider the S-dual model where the black hole is composed
of solitonic 5-branes, elementary strings, and momentum along the
elementary strings. In this case with regard to $V_4$ and $g$ dependence
\eqn\twotwo{r_1\sim [G^{(5)}L\TS]^{1/2}\sim V_4^{-1/2}g }
\eqn\twothree{r_5\sim [G^{(5)}V_4L\TS^3g^{-2}]^{1/2} \sim 1}
\eqn\twofour{r_1r_5\sim V_4^{-1/2}g }
\eqn\twonfive{\sigma^l\sim (r_1r_5)^{2l+2}\sim V_4^{-1-l}g^{2l+2} }
The dependence in \twonfive\ is the same as that in the D-brame model.

Let us postulate that when the incoming scalar is absorbed then the
initial string bound to the 5-brane splits into a total of $l+1$ strings, all bound
 to the 5-brane. (Thus for the case $l=0$ we 
have just one string in the final state, as was the case in \dasmathurone .) The 
total number of strings involved in the intertaction is $l+3$, because the initial
 state had a masslesss scalar and the initial string bound to the 5-brane. Each 
string has a normalisation factor $V_4^{-1/2}$, and the amplitude also has a factor 
$V_4$ from the volume of the interaction region. The cross section contains then 
the square of the resulting $V_4$ dependence:
\eqn\twonsix{\sigma^l\sim [V_4^{-(l+3)/2}V_4]^2\sim V_4^{-l-1} }
which agrees with \twonfive . 

Note that we have assumed here that the volume $V_4$ is small, and the wavelength of 
the incoming scalar is large, so that there is no energy to excite momentum modes of
 the  strings in the directions of the torus $T^4$. In the opposite limit, where such
 momentum modes are in fact continuous, we would have a sum
$\sum_n\sim V_4\int d^4k$ for each string in the final state, and we would  obtain 
that $\sigma^l\sim V_4^{-1}$ for all $l$, which is not in agreement with \twonfive .

Now note that the amplitude depends on $g$ as
$g^{l+1}$, for a tree vertex involving $l+3$ closed strings, which gives in the 
 cross section
\eqn\twonsix{\sigma^l\sim g^{2l+2} }
which also agrees with \twonfive . Thus we get the powers of both $V_4$ and $g$ 
to agree at the same time, which a priori need not have been the case.

\subsec{Spin dependence.}

Let us see how the strings on the 5-brane world volume carry the
angular momentum of the 4+1 dimensional transverse space. The rotation group
of the transverse space is $SO(4)=SU(2)\times SU(2)$. The string is confined
to the 5-brane, and its low energy bosonic excitations  are thus
vibrations in the compact directions, labelled by an index $i=6,7,8,9$ for
the 4 directions in the 5-brane transverse to the string. If we
quantise the string by an NSR prescription, we would take fermions
 $\psi^i, i=6,7,8,9$, and it is not immediately clear where the
angular behavior in the directions $X^1, X^2, X^3, X^4$ would come from.

But we can rigorously prove that the ground states of a string bound to a 
5-brane can carry spin for the directions $X^1, X^2, X^3, X^4$. By a sequence of
S dualitities and T dualities in the compact directions, we can map the
D 5-brane bound to a D-string to a D-string carrying say a left 
moving momentum mode. The 5 brane has been transformed to the D-string, and the
D-string has been transformed to the momentum mode. But the momentum mode
can be one of 8 bosonic and 8 fermionic states, which were described in \dasmathur .
 Out of the bosonic modes, 4 are in directions perpendicular to the
compact directions, so we see that there should be a bosonic vector state
 of the transverse $SO(4)$ among the ground states of the D-string bound to
 the D-5-brane. Similarily we find that the fermions are spinors of one
of the two $SU(2)$ components of this transverse $SO(4)$. 

An analysis of the spin properties of the string bound to
the 5-brane can be found in \ref\dvvone{R. Dijkgraaf, E. Verlinde and 
H. Verlinde, hepth 9603126.}. (Since the string \dvvone\ was quantised while
 ignoring the fact that it was not a critical string, one may argue that 
this is not a rigorously correct deriuvation of the degrees of freedom. 
But the argument of duality in the above paragraph shows rigorously that
 the obtained spin properties of the ground state are correct.) 
There is indeed a 
bosonic vector state, and fermionic states that are spinors, for the
transvese $SO(4)$. Thus the ground states can be writen as
\eqn\fivepone{(|\alpha, k_L>\oplus |a, k_L>)\otimes
(|\dot\beta, k_R>\oplus |b, k_R>)  }
Here $|\alpha, k_L>$ are the two left moving bosonic ground states, while
 $|a, k_L>$ are the two left moving fermionic ground states.
Overall we get 16 ground states, of which 8 are bosonic and 8 are fermionic,
 just as expected from the above argument through duality.

The vibrations of the string are the following. There are bosonic modes
 $X^i, i=6,7,8,9$ that can travel left or right on the string.
There are fermionic modes $\lambda^\alpha_a$ which travel left on the string and $\lambda^{\dot\beta}_b$ that travel right on the string. Note that the ground
 states of the string on the left side, say,  carry either the index $a$ or 
the index $\alpha$, while the travelling fermionic modes carry two indices $\alpha, a$.

To get the spin required of the final state, we postulate that each time a new
 closed string is produced by splitting, one left and one right moving fermion
 wave is produced on the initial string. The new closed string is taken to be
 in its ground state, with polarisations given by $(a, \dot a)$ so that there
 is no spin of the transverse $SO(4)$ carried by this string. One possible 
form of the interaction, for the case $l=1$, is
\eqn\twoneight{p^i\gamma^i_{\alpha\dot\beta}
\lambda^\alpha_a\lambda^{\dot\beta}_b\lambda'_a\lambda'_b }
where the  $\lambda'$ refer to the polarisations of the new 
closed string in its ground state, and the $\lambda$ are the
 fermionic waves that are created on the (long) initial string during
the process of absorption of the scalar. (The spatial momentum $p^i$ of the incident
 scalar
has components only in the transverse directions, since we are considering
neutral scalars.)

The details of such interactions are, however, not clear. In particular normalisation 
factors suggest that the new strings that are produced will have small winding number, 
also such strings will prefer to be in their ground states because they cannot support
 very low energy excitations. Summing over winding numbers may give rise to additional
 logarithms, not present in the classical cross section
\oneonez ,\oneonezp .

\newsec{The $\omega$ dependence.}

\subsec{Sources of $\omega$ dependence.}

We find $\omega$ dependence of the absorption cross section from the following sources:

(a)\quad As explained in \cgkt , the absorption cross section is not given by 
$\Gamma(\omega)$, the absorption when unit flux is incident, but by
\eqn\twone{\sigma(\omega)=\Gamma(\omega)-\Gamma(-\omega) }
The reason is that while the systen can absorb from the incident flux, it can 
also radiate at the same time, and the absorption cross section only measures 
the net amount of absorption. Thus we will apply \twone\ to find $\sigma(\omega)$
 after computing $\Gamma(\omega)$; the steps below pertain to the calculation of 
the latter quantity.

(b)\quad The amplitude contains a factor of the energy
$|\omega_1|$  of each boson that is produced, and also the normalisation factor
 for the boson which is
$\omega_1^{-1/2}$. So we get a factor $\omega_1^{1/2}$ in the amplitude for each
 boson, and thus a factor $|\omega_1|$ in the cross section. There is one left
 moving and one right moving boson.

(c)\quad The incoming scalar contributes a normalisation factor $\omega^{-1/2}$ 
in the amplitude, which gives $\omega^{-1}$ in the cross section.

(d)\quad The excitations on the string have a left temperature $T_L=\beta_L^{-1}$
 and a right temperature $T_R=\beta_R^{-1}$.
The incoming quantum interacts with the string through a vertex that involves one 
boson and $l$ fermions on each side, when the angular momentum absorbed is $l$. 
These bosons and fermions can either be added to the initial state of the string
 or can be absorbed from the initial state. The analysis of weight factors for 
these two cases was carried out for bosonic excitations in \cgkt . We repeat 
such an analysis for our case here.

Consider either the left or the right set of variables, and let the inverse
 temperature be denoted by $\beta$. The distribution function for bosons is $\rho_B=(e^{\beta\omega}-1)^{-1}$ and for fermions is 
$\rho_F=(e^{\beta\omega}+1)^{-1}$. 
If the boson appears in the final state then $\omega>0$ and the weightage
 factor is
$1+\rho_B(\omega)=-\rho_B(-\omega)$. If the boson was absorbed from the 
initial state
 then $\omega<0$ and the weightage factor is
$\rho_B(-\omega)$. Note that the weight factor from part (b) above is 
always positive;
 as a consequence the two cases of the boson being in the final and in the intial
 state can be combined to
have an integral 
\eqn\twtwo{-\int_{-\infty}^\infty d\omega \omega\rho_B(-\omega)}
Similarily, a fermion in the final state has $\omega>0$ and a weight $1+\rho_F(\omega)=\rho_F(-\omega)$. A fermion in the initial state has $\omega<0$
 and
a weight $\rho_F(-\omega)$. The two cases can thus be combined to an integral
\eqn\twthree{\int_{-\infty}^\infty d\omega \rho_F(-\omega) }

(e)\quad We have on each of the left and right sides the energy conservation delta
 function 
\eqn\twfour{\delta({\omega\over 2}-\sum_{i=1}^{l+1}\omega_i) }
where $\omega_1$ is the energy of the boson and $\omega_i, i=2\dots l+1$ are the
 energies of the fermions. Note that we are considering the absorption of 
neutral quanta, so
 half the energy $\omega$ goes to left movers and half to right movers.

(f)\quad We assume that the absorption vertex has a factor
$\omega/2$ for each pair of fermions (one left and one right) that are involved 
in the interaction. The factor $1/2$ is added for convenience; since we are not
 computing the actual numerical amplitude here it is of no real significance. 
But the factor $\omega$ can be seen in the disc amplitude. In a Green-Schwarz 
formalism, the fermions either have the form $\sim S$, or 
the form $\sim (\partial X) S$. The term $\partial X$ is contracted with the 
factor $e^{ikX}$ in the incident scalar vertex, and gives a factor $|k|=\omega$.
 The absorption of angular momentum $l$ needs
$l$ fermion vertices of each type, giving $(\omega/2)^l$ in the amplitude, and thus 
$(\omega/2)^{2l}$ in the cross section.

(g)\quad When we integrate over the energies of the $l$ fermions on each of the
 left and right sides, we overcount possibilities because the fermions are
 actually indistinguishable. Thus we must
correct by a factor $[l!]^{-2}$. In more detail, we note that if we consider
 the fermions travelling on a D-string bound to a 5-D-brane, then these fermions
 carry spin indices which may distinguish them. We can choose coordinates and 
consider the absorbtion of  a suitable partial wave such that all the fermions
 have the same spin in the transverse $SU(2)\times SU(2)$. But the left fermions
 also carry an index $a=1,2$ for
the spin within the directions of the 5-D-brane.  (The right fermions similarily
 carry an index $\dot a=1,2$.) Thus for the left fermions we have to symmetrise 
the fermions with $a=1$ amomg themselves and the  fermions with $a=2$ among 
themselves.
Thus we get the sum
\eqn\tweight{\sum_{j=0}^l {1\over j!}{1\over (l-j)!}
={1\over l!}\sum_{j=0}^l {}^lC_j = {2^l\over l!} }
so that we still get the desired factorial, with the extra freedom of the  
index $a$ 
giving the factor $2^l$.

\subsec{Example: $l=1$}

We have one boson and one fermion on each of the left and right sides.
 Following
 steps (b)-(g) above we obtain
\eqn\twnine{\eqalign{\omega &[\int_{-\infty}^\infty d\omega_1d\omega_2\delta(\omega/2-\omega_1-\omega_2) \omega_1 (-\rho_B^{\beta_L}(-\omega_1))\rho_F^{\beta_L}(-\omega_2)]\cr
&[\int_{-\infty}^\infty d\omega_1d\omega_2\delta
(\omega/2-\omega_1-\omega_2) \omega_1 (-\rho_B^{\beta_R}
(-\omega_1))\rho_F^{\beta_R}(-\omega_2)]\cr} }
where we have noted the temperature dependence of the 
distribution functions.

But
\eqn\twel{\eqalign{\int_{-\infty}^\infty d\omega_1\omega_1 (-\rho_B^{\beta_L}(-\omega_1))\rho_F^{\beta_L}(\omega/2-\omega_1)&=
\int_{-\infty}^\infty d\omega_1\omega_1 {1\over 1-e^{-\beta_L\omega_1}}
{1\over 1+e^{-\beta_L(\omega/2-\omega_1)}}\cr
 =
J_{BF}(\beta_L,\omega/2)=&{(\omega^2+(2\pi T_L)^2)\over 2!2^2
(1+e^{-{\omega\over 2T_L}}) }\cr } }
where $J_{BF}$ is defined in Appendix A.

Using \twone\ we get the contribution
\eqn\twfourt{{e^{\omega/T_H}-1\over (e^{{\omega\over 2T_R}}+1)
(e^{{\omega\over 2T_L}}+1)}{\omega\over (2!)^22^4}(\omega^2+(2\pi T_L)^2)
(\omega^2+(2\pi T_R)^2)}

\subsec{Example: $l=2$}

We have one boson and two fermions on each of the left and right sides. 
Following steps (b)-(g) above we obtain
\eqn\twfift{\eqalign{\omega^3 &[\int_{-\infty}^\infty
 d\omega_1d\omega_2d\omega_3 \delta
(\omega/2-\omega_1-\omega_2-\omega_3)\omega_1 (-\rho_B^{\beta_L}(-\omega_1))\rho_F^{\beta_L}(-\omega_2)
\rho_F^{\beta_L}(-\omega_3)]\cr
&[\int_{-\infty}^\infty d\omega_1d\omega_2d\omega_3\delta
(\omega/2-\omega_1-\omega_2-\omega_3) \omega_1 (-\rho_B^{\beta_R}(-\omega_1))\rho_F^{\beta_R}(-\omega_2)
\rho_F^{\beta_R}(-\omega_3)]\cr} }

But
\eqn\twel{\eqalign{\int_{-\infty}^\infty d\omega_1d\omega_2&\omega_1 (-\rho_B^{\beta_L}(-\omega_1))\rho_F^{\beta_L}(\omega_2)\rho_F^{\beta_L}
(\omega/2-\omega_1-\omega_2)\cr
&=
\int_{-\infty}^\infty d\omega_1\omega_1 {1\over 1-e^{-\beta_L\omega_1}}{1\over 1+e^{-\beta_L\omega_2}}{1\over 1+e^{-\beta_L(\omega/2-\omega_1-\omega_2)}}\cr
& =
J_{BF^2}(\beta_L,\omega/2)={\omega(\omega^2+(2\pi T_L)^2 2^2)\over 3!
2^3(1-e^{-{\omega\over 2T_L}}) }\cr } }

Using \twone\ we thus obtain the contribution
\eqn\twfourtp{{e^{\omega/T_H}-1\over (e^{{\omega\over 2T_R}}-1)
(e^{{\omega\over 2T_L}}-1)}{\omega^5\over 4(3!)^2(2!)^22^4}
(\omega^2+(2\pi T_L)^2)(\omega^2+(2\pi T_R)^2)}

\subsec{General form of $\omega$ dependence.}

For $l$ odd we obtain
\eqn\twtw{\eqalign{{1\over 4}{1\over 2^{2l}}&{e^{\omega/T_H}-1\over
 (e^{{\omega\over 2T_R}}+1)
(e^{{\omega\over 2T_L}}+1)}{\omega^{2l-1}\over (l!)^2((l+1)!)^2}\cr
&[\omega^2+(2\pi T_L)^21^2][\omega^2+(2\pi T_L)^23^2]\dots 
[\omega^2+(2\pi T_L)^2l^2]\cr
&[\omega^2+(2\pi T_R)^21^2][\omega^2+(2\pi T_R)^23^2]\dots 
[\omega^2+(2\pi T_R)^2l^2]\cr }}

For $l$ even we obtain
\eqn\twtwp{\eqalign{{1\over 4}{1\over 2^{2l}}&{e^{\omega/T_H}-1\over
 (e^{{\omega\over 2T_R}}-1)
(e^{{\omega\over 2T_L}}-1)}{\omega^{2l+1}\over (l!)^2((l+1)!)^2}\cr
&[\omega^2+(2\pi T_L)^22^2][\omega^2+(2\pi T_L)^24^2]\dots 
[\omega^2+(2\pi T_L)^2l^2]\cr
&[\omega^2+(2\pi T_R)^22^2][\omega^2+(2\pi T_R)^24^2]\dots 
[\omega^2+(2\pi T_R)^2l^2]\cr }}

We see that these dependences on $\omega$ agree with the dependences required
by the  classical cross section \oneonez , \oneonezp .

\newsec{Discussion.}

We have seen that to have the classical absorption agree with the
brane models, we need to obtain a dependence on $V_4$ (the volume
of the compact torus perpendicular to the string) in the brane model.
 While this may show up in different ways in different treatments of 
the string dynamics, it is possible that these differences are due to
 the different coupling regimes appropriate to these calculations, and 
not to an error in either description.
Recently it has been shown that there is `stringy dynamics' in all 
higher branes, in some domain of parameters \ref\lind{U. Lindstrom 
and R von Unge, hepth 9704051.}.

The issue of $V_4$ dependence may appear in other calculations, for
example that for the fixed scalars  in \cgkt . In this calculation 
it was assumed that $r_1=r_5$, which is equivalent to choosing a 
particular value for $V_4$. It would be interesting to see the
details of the agreement when $r_1\ne r_5$.

Regardless of the details of the absorption, we note that
the desired $\omega$ dependence arises from a partitioning
of the energy of the incoming scalar into the energy of
a certain number of momentum modes, with this number being determined 
by the angular momentum of the partial wave that is absorbed. The 
calculation also provides naturally the factorials
present in the relative cross sections for different $l$ (though 
since we have not computed the actual disc amplitudes themselves,
 we cannot know that there will be no other factorials from that source.) 
The argument using an `effective conformal theory' carried out
in \maldastromtwo\ yielded an $\omega$ dependence that was the desired one,
but there was no known way to normalise the amplitude. The energy 
dependence calculation of the effective
conformal theory calculation is plausibly equivalent to
summing over ways of sharing energy between $l+1$ quanta of the
left and right sides
when the quanta are at  temperatures $T_L, T_R$ respectively,
since one has to evaluate correlators of free fields at
the appropriate
temperatures. 
 
In \klebanovmathur\ it was noted that when we are working in a domain
$r_5>> r_1, r_p, r_0$ then the absorption of the $l$th partial wave 
becomes significant when $\omega r_5=l+1$. In particular the
$l=1$ partial wave is significantly absorbed starting at the energy where 
the first massive mode of the effective string can be created. Creation of
 a new string state would bring in the required factor of $V_4$ as we have
 seen. It would be interesting to connect the present calculations to this 
domain of parameters where $r_1$ is also small, and so winding modes and 
momentum modes play a more
symmetric role.

It may be thought that the agreement of cross sections for D-branes at low 
energy with
the black hole cross sections implies that for low energy quanta we understand
 the 
mechanism by which the Hawking paradox is to be resolved in string theory. 
This is not the case. A black hole geometry takes an incident  low energy
 quantum and, 
in the Schwarzschild coordinate system, converts it to a high frequency mode
 close
 to the horizon. This
high frequency mode is then eaten by the hole. Starting with a quantum of even 
lower energy
just means that we have to follow the mode closer to the horizon before
we see it attaining a short wavelength. (Of course most of the low energy quanta
escape falling into the hole, but the cross section we compute relates to those
 that
are in fact swallowed by the hole in the above fashion.)

Equivalently, if we study the deflection of a particle trajectory by the 
gravitational field
of the hole, and consider the deflection to be expanded in powers of the mass
of the hole, then we would see a divergence of the series when the impact
 parameter approaches the
value where the particle will be swallowed by the hole. When we study the
D-brane calculation we need to understand whether or not such a divergence
 occurs,
when the number of branes and the coupling are increased to a value large 
enough to
give a classical sized black hole. In the classical calculation taking low 
energy while keeping other parameters fixed simply pushes the growth of the 
perturbation series towards higher
terms in the series; thus there may be a similar phenomenon for
the D-brane calculation as well.

If string theory is to resolve the Hawking paradox, then we either need to see
that the effective size of the solitonic bound state is comparable to the horizon
size, so that there is really no black hole, or we need to see that
loops of virtual quanta in a theory with strings and higher dimensional
branes are quite different from loops in particle theory, and give nonlocal
effects that take information from near the singularity and send it
out with the Hawking radiation. The latter is equivalent to finding
a length scale in string theory that is not plank length but is a length that grows 
with
the number of  branes involved. 

Thus it appears that the agreements found between  cross sections of branes
and for black holes are to some extent both mysterious and interesting, and
provide a strong suggestion that the black hole paradox may actually
be resolved in string theory. One needs to better understand the bound
states of many branes  at strong coupling; for the non-BPS interactions
that are involved the 
values of moduli also affect the nature of energy levels and
interaction properties \ref\mathur{S.D. Mathur, hepth 9609053.}\klebanovmathur .

\bigskip
\bigskip
\bigskip


\centerline{{\bf Acknowledgements}}

I would like to thank S.P. deAlwis, K. Johnson,  I. Klebanov and 
G. Lifschytz for discussions. I am especially
grateful to S.R. Das for discussions and a  critical reading of the
 manuscript.
This work is partially supported by cooperative
agreement number DE-FC02-94ER40818.
 
\vfill
\eject

\appendix{A}{The basic integrals.}

Let us define the following two basic integrals
\eqn\sixone{I_{BF}^n\equiv
\int_{-\infty}^\infty {x^ndx\over (e^x-1)(e^{-x}+A) } }
\eqn\sixonep{I_{FF}^n\equiv
\int_{-\infty}^\infty {x^ndx\over (e^x+1)(e^{-x}+A) } }
(The subscripts $B$ and $F$ stand for bose and fermi type distributions
 respectively.)

These integrals can be calculated in the following way. We take a contour
 in the complex $x$ plane, running along the real $x$-axis,
and backwards along the line $x+2\pi i$. 
Consider the integral
\eqn\sixthree{\hat I_{BF}^n\equiv
\int_C {x^n(x-2\pi i)dx\over (e^x-1)(e^{-x}+A) } } 
The segments at $\Re x=\pm\infty$ do not contribute. So we have
\eqn\sixthreep{\hat I_{BF}^n = 
\int_{-\infty}^\infty {[x^n(x-2\pi i)-(x+2\pi i)^nx]dx\over (e^x-1)(e^{-x}+A) } }
where now $x$ runs along the real line.
There is one pole in the contour, at $x=i\pi-\log A$. Then we find
\eqn\sixtwo{(n+1)I^n_{BF}+\sum_{m=0}^{n-2}{}^nC_mI^{m+1}_{BF}(2\pi i)^{n-m-1}=
{(\pi^2+(\log A)^2)(i\pi-\log A)^{n-1}\over A+1} }
It will turn out that we will only need $I^n_{BF}$ for $n$ odd.
From \sixtwo\ relation we find
\eqn\sixfour{I_{BF}^1={(\pi^2+(\log A)^2)\over 2 (A+1)} }
\eqn\sixfourp{I_{BF}^3={(\pi^2+(\log A)^2)^2\over 4 (A+1)} }
\eqn\sixfourpp{I_{BF}^5={(\pi^2+(\log A)^2)(3\pi^2+(\log A)^2)\over 6 (A+1)} }
More generally, for $n$ odd
\eqn\sixfourq{\eqalign{I_{BF}^n=&{(i\pi)^{n-1}(\pi^2+(\log A)^2)\over (n+1)(A+1)}
[(\log A)^{n-1}+{(n+3)(n-2)\over 6}(\log A)^{n-3}\cr
&+{(n-2)(n-4)(7n^2+28n+45)\over 360}(\log A)^{n-5}+\dots ]\cr} }

Similarily, we define
\eqn\sixthreepp{\hat I_{FF}^n\equiv
\int_C {x^{n+1}dx\over (e^x+1)(e^{-x}+A) }=\int_{-\infty}^\infty 
{[x^{n+1}-(x+2\pi i)^{n+1}]dx\over (e^x+1)(e^{-x}+A) }  } 
where the contour $C$ is the same as described above, and in the last integral
 $x$ runs  over the real line.
There are two poles, at $x=i\pi$ and at $x=i\pi-\log A$.
We get the relation
\eqn\sixsix{\sum_{m=0}^n {}^{n+1}C_mI^m_{FF}(2\pi i)^{n-m}=
-{1\over A-1}[(i\pi-\log A)^{n+1}-(i\pi)^{n+1}] }
We will need $I_{FF}^n$ only for even $n$. We find
\eqn\sixfive{I^0_{FF}={\log A\over A-1} }
\eqn\sixeight{I_{FF}^2={\log A(\pi^2+(\log A)^2)\over 3(A-1)} }
\eqn\sixeightp{I_{FF}^4={\log A(\pi^2+(\log A)^2)
({7\over 3}\pi^2+(\log A)^2)\over 5(A-1)} }
More generally, for $n$ even
\eqn\sixfourqq{\eqalign{I_{FF}^n=&{(i\pi)^{n}(\pi^2+(\log A)^2)\over (n+1)(A-1)}
[(\log A)^{n}+{n(n+1)\over 6}(\log A)^{n-2}\cr
&+(n+1)n(n-1)(n-2){7\over 360}(\log A)^{n-4}+\dots ]\cr} }

We now define the integrals $J_{BF^n}(\beta, \omega)$. The arguments of
$J_{BF^n}$ will not be explicitly written below, for convenience. 
(We also write $J_{BF}\equiv J_{BF^1}$.)

The integral $J_{BF}$ arises if we wish to partition energy $\omega$
 between one boson and one fermion. The integral is
\eqn\sixel{J_{BF}=-\int_{-\infty}^\infty d\omega_1 \omega_1
\rho_B(-\omega_1)\rho_F(-\omega_2)\delta(\omega_1+\omega_2-\omega) =
 -\int_{-\infty}^\infty {d\omega_1 \omega_1 \over (e^{-\beta\omega_1}-1)
(e^{-\beta(\omega-\omega_1)}+1 )
 } }
Define $x=-\beta\omega_1$. Then we see that
\eqn\sixtw{J_{BF}={1\over \beta^2e^{-\beta\omega}}\int_{-\infty}^\infty 
{xdx \over (e^x-1)(e^{-x}+A)} }
where
\eqn\sixthir{A=e^{\beta\omega} }
But from \sixfour\ 
\eqn\sixnone{J_{BF}={1\over \beta^2e^{-\beta\omega} }I^1_{BF}=
{(\omega^2+{\pi^2\over \beta^2})\over 2!(1+e^{-\beta\omega}) } }

Similarily, suppose we wish to partition the energy $\omega$ between
 one boson and two fermions. Then the integral we obtain is
\eqn\sixtw{J_{BF}^2=\int_{-\infty}^\infty{\omega_1d\omega_1d\omega_2\over (1-e^{-\beta\omega_1})(1+e^{-\beta\omega_2})
(1+e^{-\beta(\omega-\omega_1-\omega_2)})}}
First we compute
\eqn\sixtwone{\int_{-\infty}^\infty{\omega_1d\omega_1\over (1-e^{-\beta\omega_1})(1+e^{-\beta(\omega'-\omega_1)})}
={1\over \beta^2}I_{BF}^1(e^{\beta\omega'})=
{1\over \beta^2}{(\pi^2+(\omega'\beta)^2)\over 2(1+e^{-\beta\omega'})}} 

Then we find
\eqn\sixtwtwo{J_{BF}^2={1\over 2\beta^3}[\pi^2I_{FF}^0(e^{\beta\omega})+\beta^2I_{FF}^2(e^{\beta\omega})]=
{\omega\over 6(e^{\beta\omega}-1)}
(4{\pi^2\over\beta^2}+\omega^2) }

For the energy to be shared between one boson and 3 fermions, we get
\eqn\sixtwthree{J_{BF^3}={1\over 6}[4{\pi^2\over\beta^2}{I_{BF}^1
(e^{\beta\omega})\over \beta^2}
+{I_{BF}^3(e^{\beta\omega})\over \beta^4}]={({\pi^2\over\beta^2}+\omega^2)(9{\pi^2\over\beta^2}+\omega^2)
\over 4! (e^{\beta\omega}+1)}}

Similarily we find
\eqn\sixtwfour{J_{BF^4}={\omega\over 5!(e^{\beta\omega}-1)}
(4{\pi^2\over\beta^2}+\omega^2)(16{\pi^2\over\beta^2}+\omega^2)}
\eqn\sixtwfourp{J_{BF^5}={1\over 6!(e^{\beta\omega}+1)}
({\pi^2\over\beta^2}+\omega^2)(9{\pi^2\over\beta^2}+\omega^2)
(25{\pi^2\over\beta^2}+\omega^2)}
\eqn\sixtwfourp{J_{BF^6}={\omega\over 7!(e^{\beta\omega}-1)}
(4{\pi^2\over\beta^2}+\omega^2)(16{\pi^2\over\beta^2}+\omega^2)
(36{\pi^2\over\beta^2}+\omega^2)}

While we have not solved for the general term $J_{BF^n}$ these cases provide a 
pattern that we 
will assume holds for all $n$. In particular we note the appearance of the factor
$1/(n+1)!$ in the integrals.

\appendix{B}{Angular variables.}

We have the transverse rotation group $SO(4)=SU(2)\times SU(2)$. Let the
generators of $SO(4)$ be $M_{ij}, i,j=1\dots 4$. Then the generators for
the two $SU(2)$ components are described through
\eqn\twone{\eqalign{J_1=&{1\over 2}(M_{12}+M_{34}) \cr
J_2=&{1\over 2}(M_{13}+M_{42}) \cr
J_3=&{1\over 2}(M_{32}+M_{41}) \cr}}
and
\eqn\twonep{\eqalign{K_1=&{1\over 2}(M_{12}-M_{34}) \cr
K_2=&{1\over 2}(M_{13}-M_{42}) \cr
K_3=&{1\over 2}(M_{32}-M_{41}) \cr}}

We have
\eqn\twtwo{[J_a,J_b]=\epsilon_{abc}J_c, ~~[K_a,K_b]=\epsilon_{abc}K_c,
 ~~[J_a,K_b]=0}

The spinor of $SO(4)$ is contructed as $(1/2,0) \oplus (0,1/2)$. Letting
the spinor components in each $SU(2)$ be represented by $\{+,-\}$, we have the
four components $\{++, +-, -+, --\}$. The rotation generators are
$M_{ij}={1\over 4}[\Gamma_i,\Gamma_j]$. This gives
\eqn\twthree{\eqalign{{1\over 2}(M_{12}\pm M_{34})=&{i\over 4}(\sigma^3\otimes 
1\pm 1\otimes\sigma^3)\cr
{1\over 2}(M_{13}\pm M_{42})=&-{i\over 4}(\sigma^2\otimes 
\sigma^1\pm \sigma^1\otimes\sigma^2)\cr
{1\over 2}(M_{32}\pm M_{41})=&-{i\over 4}(\sigma^1\otimes 
\sigma^1\pm \sigma^2\otimes\sigma^2)\cr }}

The vector of $SO(4)$ decomposes as $(1/2,1/2)$ under
the two $SU(2)$ components. Let us write the following combinations of
the four components of the vector
\eqn\twfour{\eqalign{X^1\pm iX^2=&X_A^\pm\cr
X^3\pm iX^4=&X_B^\pm\cr}}
Then we have the identifications
\eqn\twfive{\eqalign{X_A^+=&(++)(+-)\cr
X_A^-=&(--)(-+)\cr
X_B^+=&(++)(-+)\cr
X_B^-=&(--)(+-)\cr }}

If we tensor together two vectors we have $(1/2,1/2)\otimes (1/2,1/2)
=(0,1)\oplus (1,1) \oplus (1,0)$. The component $(1,1)$ we identify with
the  tensors $C_{ij}X^iX^j$ with $C_{ij}$ symmetric and traceless.
Thus this $(1,1)$ component is all that we expect to pick up in the absorption of
higher partial waves of scalars.

More generally, we identify the component $(l/2,l/2)$ with the partial wave
components for angular momentum $l$. The former description is seen to have 
$(l+1)^2$ components. To see that the latter also has the same number of components,
note that the number of symmetric tensors that can be made from $l$
copies of the vector $X^i$ is ${}^{(l+3)}C_3=(l+3)(l+2)(l+1)/6$. We
must remove all those components that can arise when any two of the $X^i$
components are traced over, and this gives ${}^{(l+1)}C_3=(l+1)(l)(l-1)/6$
components. The difference is $(l+1)^2$, as required.

Thus in absorbing the $l$th partial wave of the scalar we will create
spins on the brane system that total to a value $l/2$ in each of the $SU(2)$
components.

\appendix{C}{Relating plane waves to partial wave components.}

We wish to find the partial wave components in a plane wave. In
3 space dimensions, we have the decomposition
\eqn\twsix{e^{i\omega z}=e^{i\omega r\cos\theta}\rightarrow
\sum_{l\ge 0}{e^{-i\omega r}\over (-i\omega r) }(-1)^l[\pi(2l+1)]^{1/2}Y_{l,0}
(\cos\theta) }
where $Y_{l,0}$ is a function of the angular coordinates only and is
 normalised to satisfy $\int |Y_{l,0}|^2d\Omega=1$. Thus if the
absorption probability of the $l$th partial wave is $a_l$, then the cross section
for this partial wave is
\eqn\twseven{\sigma^l={\pi\over \omega^2}(2l+1)a_l}
Note in particular that the conversion factor from $a_l$ to $\sigma^l$ does have 
 a factor $2l+1$ even though only one azimuthal component is seen in \twsix . One
 way to understand the  factor $2l+1$ is to let the incident wave
be averaged over all directions, at which stage we still expect 
the same cross section since we have just
done an averaging. But now we can regard the incoming quanta as
 decomposed into partial waves, and study the radial motion of the wave as a problem
 in
$1+1$ dimensions. There are $2l+1$ `species' of radially incoming quanta,
and so the cross section has  a factor $2l+1$.

In 4 space dimenions we have the decomposition
\eqn\twsixp{e^{i\omega X^4}=e^{i\omega r\cos\theta}\rightarrow
\sum_{l\ge 0}{e^{-i\omega r}\over (\omega r)^{3/2}}e^{i3\pi/4}(-1)^l
\sqrt{4\pi}(l+1) Z_{l,0}(\cos\theta) }
where
\eqn\tweleven{Z_{l,0}(\cos\theta)={1\over \sqrt{2\pi^2}}U_l(\cos\theta)
={1\over \sqrt{2\pi^2}}{\sin[(l+1)\theta]\over
\sin\theta} }
where the $U_l$ are Chebyshev polynomials. The $Z_{l,0}$ are normalised according
 to
\eqn\twnine{\int |Z_{l,0}|^2d\Omega=\int_0^\pi |Z_{l,0}|^2 4\pi\sin^2\theta
d\theta=1}

The cross section for absorption of the $l$th partial wave is given in terms
of its absorption probability $a_l$ through
\eqn\twseven{\sigma^l={4\pi\over \omega^3}(l+1)^2a_l}
Similar to the case of 3 space dimensions, the number
$(l+1)^2$ is the number of azimuthal components for angular momentum $l$, as was
noted above.

\listrefs

\end